\documentclass{article}%
\usepackage{amsmath}
\usepackage{amsfonts}
\usepackage{amssymb}
\usepackage{graphicx}
\usepackage{multirow}

\title{H\'{e}non's generating solutions and the structure of periodic orbits
families of the restricted three-body problem}
\author{Alexander B. Batkhin\\
Keldysh Institute of Applied Mathematics, Moscow, Russia\\
e-mail: batkhin@gmail.com\\
Natalia V. Batkhina\\
Volgograd State University, Volgograd, Russia\\
e-mail: nbatkhina@gmail.com\\
}

\newcommand {\eps}{\varepsilon}

\DeclareMathOperator{\Hess}{Hess}

\newtheorem{definition}{Definition}
\newtheorem{proposition}{Proposition}

\begin{document}
\maketitle

\vspace*{-1cm}

\begin{abstract}
We propose a survey of Michel H\'enon works devoted to studying periodic solutions of the well-known celestial mechanics problem -- restricted three-body problem. The description of the main results obtained by H\'enon is given in comparison with results of russian mathematician Alexander Bruno. 
Finaly, a survey of H\'enon's works on Hill problem is given as well, and, more over, the authors propose some generalization of Hill problem that makes possible to provide the description of its families of periodic orbits in the form of common network.



\end{abstract}

\section{Introduction}

There are two problems in theoretical mechanics, which influence on developing of many branches of mathematics and mechanics can not be underestimated:
\begin{itemize}
\item the problem of rotation of rigid body  with one fixed point;
\item the restricted three-body problem (RTBP).
\end{itemize}

Planar circular RTBP is an example of rather simple but non-integrable Hamiltonian system with two degrees of freedom, which possesses all the richness of its phase flow behavior. The most interesting objects of this problem are periodic orbits and their families. On our opinion the following quote from~\cite[Ch.~III, \S~36]{PoincareI} gives us an explanation of importance of periodic orbits.

\begin{quote}
{\ldots ce qui nous rend ces solutions p\'eriodiques si pr\'ecieuses, c'est qu'elles sont, pour ainsi dire, la seule br\'eche par o\`u nous puissions essayer de p\'en\'etrer dans une place jusqu'ici r\'eput\'ee inabordable.}\\
\end{quote}

Families of periodic solutions of dynamic systems with two degrees of freedom form a frame of system behavior. That is why a lot of methods of periodic orbits calculation  were developed, but overwhelming majority of these methods are designed for so called regular cases, when solution of unperturbed system moves far away of singular points of perturbing function. The singular case is much more complicated and regular case methods do not applicable. Michel H\'enon provided an approach based on a concept of so called ``generating solution''. This approach together with the highest level of numerical computations allowed to reach the understanding of structure of the main families of periodic solutions of the planar circular RTBP.

The first H\'enon's  paper~\cite{HenonI} devoted to periodic solutions of the RTBP was published in 1965 and the last his paper~\cite{Henon2005} was published in 2005. During this period M.\,H\'enon consequently developed and used an approach based on generating solution technique for studying periodic orbits of the second species or \textit{orbits with consecutive collisions} as they were called by Poincar\'e. Almost at the same time another researcher -- professor A.\,D.\,Bruno -- provided his own studying of the RTBP's periodic solutions. They independently obtained the structure of generating solutions of the RTBP and from here mutually interacted with each other during about 30 years. In presented contribution we want to show that there was such a kind of scientific competition between Michel H\'enon and Alexander Bruno.

\section{Survey of M.\,H\'enon's works on the RTBP}
As H\'enon mentioned in the Preface to the book~\cite{Henon97} his work over generating solutions of the RTBP lasted about thirty years. During this period he interrupted his study and then came back again. May be it is possible to divide the thirty years long interval of the RTBP's orbits investigation into three main blocks.

\subsection{First block of H\'enon's study}

During 1965-70 H\'enon presented some papers devoted to numerical exploration of periodic solutions of the RTBP. 
Symmetric orbits for  ($\mu=1/2$) were investigated.

Let us recall the setting of the RTBP.

Two massive pointlike bodies $M_1$ and $M_2$ move in a plane under the influence of Newton's gravitational law around their center of mass along circular orbits. In the same plane of motion there is another one massless body $M_3$ which motion one want to study. The units of distance, mass and time are selected in a such way that the distance between $M_1$ and $M_2$, total mass of these bodies, gravitational constant are equal to unity and the period of revolution is equal to $2\pi$. In synodical i.e. uniformly rotating frame body $M_1$ is situated in the origin and the position of the body $M_2$  is point with coordinates $(1,0)$. The only parameter of the problem is so called \textit{mass parameter} $\mu$ -- the mass of the second body. The Hamiltonian function of the RTBP in rotating frame is following
\begin{equation}
H=\frac12\left(y_1^2+y_2^2\right)+x_2y_1-x_1y_2-\frac1{r_1}+\mu\left(\frac{1}{r_1}+x_1-\frac{1}{r_2}\right),
\end{equation}
where $r_1^2=x_1^2+x_2^2$, $r_2^2=(x_1-1)^2+x_2^2$.

From the practical point of view the most interesting case for investigation is when parameter $\mu$ is rather small but particular case $\mu=1/2$ -- so called Copenhagen case -- attracted its attention and H\'enon numerically studied this case as well. We guess that this case was selected by H\'enon for investigation due to presence of great computation work provided by the group of Str\"omgren in 1930-th. So he had possibility to compare his own numerical results with those that were calculated earlier and by different methods.

%
Two papers~\cite{HenonI, HenonII} demonstrated the highest level of numerical investigation of the problem. In brief the essential properties of H\'enon's work are following.
\begin{itemize}
\item Accurate and thorough numerical computations in regular coordinates using global Thiele regularization (see~\cite{Szebehelyeng}).
\item Poincar\'e map technique was used for bifurcation analysis of periodic solutions.
\item Recomputation of earlier found orbits and investigation of six new families of periodic solutions with asymptotic behavior.
\end{itemize}
Very interesting class of orbits with asymptotic behavior were studied. These orbits exist for definite values of parameter $\mu$, namely, when the eigenvalues of matrix $\Hess H$ computed at the triangular Lagrange libration points $L_4$ or $L_5$ are complex with non-zero real and imaginary parts.

The next paper~\cite{Henon68} from the first block presented the main ideas for studying second species periodic solutions with the help of generating solution technique. Let us give some definition.

\begin{definition}\label{defGenSol}
Let at small parameter value $\mu>0$ there exists a periodic solution $z(t,\mu)$ to canonical system defined by Hamitonian 
\[
H(z)=H_0(z)+\mu R(z),
\]
and it could be smoothly continued over $\mu$. Then its limit at $\mu\to0$ (if it does exist) is called \textbf{generating solution}. Here $z=(x,y)$, $x$ is a vector of coordinates and $y$ is a vector of canonically conjugated momenta.
\end{definition}

We assume that the system of canonical equations has a periodic solution $ z(t, z_0)$ with period $T$: $ z(T, z_0)= z_0$ for definite value of small parameter $\mu$. Let $\rho'_\mu$ and $\rho''_\mu$ are the minimal and the maximal distances correspondingly from the body $M_2$ to the orbit $ x(t, z_0)$. 
There are three possible ways of continuation of periodic solution $ z(t, z_0)$ while $\mu\to0$:
\begin{enumerate}
\item If $\lim\limits_{\mu\to0}\rho'_\mu>0$ then we get generating solution of the \textit{first species}; such solution is regular and can be find by the normal form method (see~\cite[Ch. II, VII]{BrunoRTBPeng}).
\item If $\lim\limits_{\mu\to0}\rho'_\mu=0$ and $\lim\limits_{\mu\to\infty}\rho''_\mu>0$ then we get generating solution of the \textit{second species}.
\item If $\lim\limits_{\mu\to0}\rho'_\mu=\lim\limits_{\mu\to\infty}\rho''_\mu=0$ then we get generating solution of the \textit{third species}.
\end{enumerate}

If generating solution is not regular then it may consist of solutions of special form called \textit{arc-solution}. The arc-solutions  start and finish at the singular points of the Hamiltonian $H$. So, generating solution of the second species can be constructed from the arc-solutions.

\begin{figure}[htb]
\centering
\includegraphics[width=.5\linewidth]{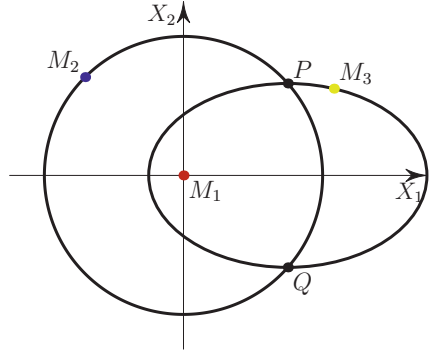}
\caption{Orbits in sidereal frame for the case $\mu=0$.}\label{figmu0}
\end{figure}

M.\,H\'enon found out two types of arc-solutions called $S$-arcs and $T$-arcs. Let consider the case $\mu=0$ in sideral i.e. inertial frame. The bodies $M_2$ and $M_3$ execute Keplerian motion, $M_2$ uniformly moves along the unit circle and $M_3$ moves along the ellipse (see Fig.~\ref{figmu0}). 

The main difference between the Kepler problem and this case is collisions between bodies $M_2$ and $M_3$. The position of the bodies are defined as follows:
\begin{equation*}
\begin{aligned}
M_2&:& x_1&=\cos\tau,& x_2&=\sin\tau,& &\\
M_3&:& x_1&=\varepsilon a\left(\varepsilon''\cos\eta-e\right),& x_2&=\varepsilon\varepsilon' a\sqrt{1-e^2}\sin\eta,& t&=a^{3/2}\left(\eta-\varepsilon''e\sin\eta\right),
\end{aligned}
\end{equation*}
where $\eta$ is true anomaly, $a$ and $e$ are the semi-major axis and the eccentricity of the ellipse, parameters $\varepsilon$, $\varepsilon'$, $\varepsilon''$ can take values $\pm1$, defining the direction of the motion and position of the pericenter of the ellipse.

Two cases of collisions are possible.
\begin{enumerate}
\item The collision takes place at the same point $P$ at the moment $t_1$. Therefore, the sideral periods of motion are commensurable and such arcs were denoted by $T$.
\item The first collision take place at the point $P$ at the moment $t_1$ and the second collision takes place at the symmetrical point $Q$ at the moment $t_2$. The set of symmetric arc-solutions was denoted by H\'enon as $S$-arcs.
\end{enumerate}

\begin{figure}[htb]
\centering
\includegraphics[width=.6\textwidth]{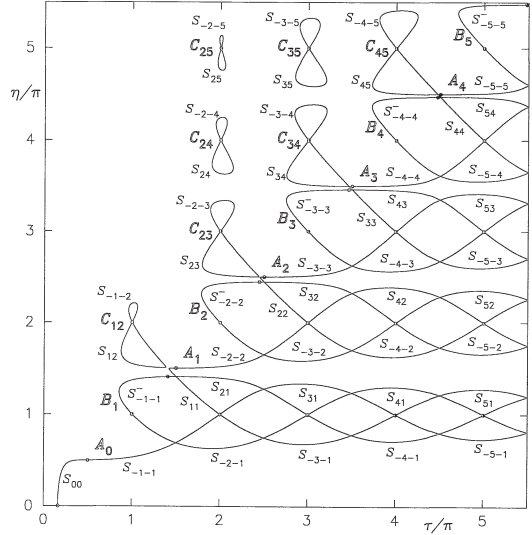}
\caption{Characteristic of arc families in $(\tau,\eta)$ plane.}\label{figHenonDiag}
\end{figure}

H\'enon obtained an equation relating time $\tau$ and true anomaly $\eta$ only
\begin{equation}\label{eqHeneq}
\begin{split}
\sqrt{1-\eps\eps''\cos\tau\cos\eta}\left[\eta(1-\eps\eps''\cos\tau\cos\eta)-\sin\eta(\cos\eta-\eps\eps''\cos\tau)
\right]-\\
-\tau|\sin\eta|^3=0
\end{split}
\end{equation}
and numerically solved it for $0<\tau<5.5\pi$ and $0<\eta<6.5\pi$. He found countable many analytic solutions of the equation~\eqref{eqHeneq} grouped into three sequences $A_i$, $B_j$ and $C_{ik}$, where indices $i\in\mathbb N_0$, $j,k\in\mathbb N$. The corresponding curves are shown in Fig.~\ref{figHenonDiag}
 taken from~\cite{Henon97}.
H\'enon also compute critical orbits of these families. This was the first step to understanding the structure of the generating solutions of the RTBP's periodic orbits.

Last two papers~\cite{HenonV, HenonVI} written at the same time were devoted to studying Hill problem which is the limiting case of the RTBP. We consider H\'enon's work on Hill problem in the next section.

\subsection{Bruno results on the RTBP}
Since the beginning of 1970th Alexander Bruno started his own program of investigation of periodic solutions of the RTBP. His candidate and doctoral theses were devoted to the problem of convergence of normal forms of ordinary differential equations and their systems. So, using this approach A.\,Bruno independently obtained the description of generating solutions of periodic orbits of the RTBP. He published his results in Soviet scientific journals and preprints of Keldysh Institute of Applied Mathematics in Russian. As H\'enon wrote in the Preface to his book~\cite{Henon97} ``I was so impressed by his work that \ldots I translated two of his papers into English \ldots''.  Finally Bruno's papers have been collected into book published in Russian in 1990 and translated into English in 1994. This book provided theoretical background of investigation of Hamiltonian systems with the method of normal form and proposed a survey of generating solutions of the first species (regular case) and of the second species (singular case). Bruno also studied bifurcations of generating solutions of the first species but the same analysis for generating solutions of the second species was not provided yet.

The direct face-to-face encounter between Michel H\'enon and Alexander Bruno took place at University of Nice at August, 1990, when Bruno was invited as visit professor.
\begin{figure}[htb]
\centering
\includegraphics[width=.7\textwidth]{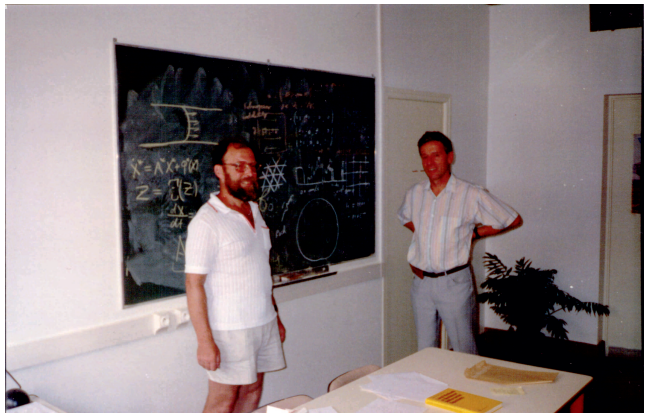}
\caption{Michel H\'enon (right) and Alexander Bruno (left) in University of Nice, August, 1990}\label{figHenonBruno}
\end{figure}

\subsection{H\'enon's books}
The second period of H\'enon's investigation of the RTBP periodic solutions was summarized in his two monographs published in Springer. The first one named ``Generating Families in Restricted Three-Body Problem'' was based on the sequel of his papers and also papers written in collaboration with Pierre Guillaume, Donald Hitzl. He also used the results obtained by Lawrence Perko and especially by Alexander Bruno. The main achievements of this book  were precise description of structure of second species generating solutions, their bifurcation analysis and comparison with so called natural families which were computed by Broucke in 1968 in the Earth-Moon case ($\mu=0.012155$) and by Bruno in 1993, 1996 in the Sun-Jupiter case ($\mu=0.00095388$). As concerning the structure of the generating solutions the main proposition was presented in Chapter 4 of the book:
\begin{proposition}[{Proposition~4.3.2 from~\cite{Henon97}}]\label{prop1}
An ordinary generating orbit of the second species cannot contain two identical $T$-arcs of type 1 in succession.
\end{proposition}

The background of this Proposition~\ref{prop1} was provided rather heuristic, so, to strictly complete generating solutions investigation and to give quantitative analysis of bifurcations, H\'enon decided to write one more book. As he wrote at the beginning of Chapter~15 in~\cite{HenonBook2} ``After the present monograph was completed \ldots there appear a book by Bruno (1998, 2000)''. Indeed, in 1998 Alexander Bruno published his new book ``Power Geometry in Algebraic and Differential Equations'' in Russian, which was translated into English and published in 2000 by Elsevier~\cite{Bruno_stepgeomeng}. The book of Bruno proposed the method for quantitative analysis of solutions near singularities and therefore gave an instrument for bifurcation analysis of generating solutions. M.\,H\'enon considerably rewrote his second volume and published its new version with strict background of bifurcation analysis of generating families in 2001.

But competition between H\'enon and Bruno did not finish. During 2005--2012 A.\,Bruno mainly together with V.\,Varin published a sequel of papers where many families of periodic solutions were computed and described in the case $\mu\to0$. Here is an abstraction from Alexander Bruno paper list of last papers, which are related to the problem of computation of families of periodic orbits with the help of their generating solutions.
\begin{enumerate}
\item[65]  On periodic flybys  of  the moon. Celestial  Mechanics  24:3 (1981) 255-268.
\item[258] Periodic solutions to a Hamiltonian system // 
Cosmic Research 44:3 (2006) 245-257.
\item[271]  On families of periodic solutions of the restricted three-body  problem (with V.P. Varin) // Celestial Mechanics and Dynamical Astronomy, 2006, v. 95, p. 27-54.
\item[283] Periodic solutions of the restricted three-body problem for small mass ratio (with V.P. Varin) // 
J. Appl. Math. Mech. 71:6 (2007) 933-960.
\item[291]  On families of periodic solutions of the restricted three-body problem (with V.P. Varin) // 
Solar System Research 42:2 (2008) 158-180.
\item[304]  Family h of periodic solutions of the restricted problem for small $\mu$ (with V.P. Varin) // 
Solar System Research 43:1 (2009) 2-25.
\item[305]  Families c and i of periodic solutions of the restricted problem for $\mu=5\cdot10^{-5}$ (with V.P. Varin) // 
Solar System Research 43:1 (2009) 26-40.
\item[306]  Family h of periodic solutions of the restricted problem for big mu (with V.P. Varin) // 
Solar System Research 43:2 (2009) 158-177.
\item[307]  Closed families of periodic solutions of the restricted problem (with V.P. Varin) // 
Solar System Research 43:3 (2009) 253-276.
\item[337]  On asteroid distribution (with V.P. Varin) // 
Solar System Research 45:4 (2011) 323-329.
\item[362] Periodic solutions of the restricted three body problem for small $\mu$ and the motion of small bodies of the Solar system (with V.P. Varin) // Astronomical and Astrophysical Transactions (AApTr), 2012, vol. 27, Issue 3, pp. 479-488.
\end{enumerate}

\section{H\'enon's works on Hill problem}
Our first acquaintance with H\'enon's papers took place in 1995 when we were offered to read two papers~\cite{HenonV, HenonVI}, devoted to Hill problem. From that moment studying families of periodic solutions of Hamiltonian systems in general and of Hill problem in particular became the main field of our scientific 
researches. At the end of 2004 when the second author was finishing her candidate thesis we decided to write Michel H\'enon a message for the purpose to get an independent evaluation of the level of fulfilled work. He asked us to send him our last results concerning non-symmetric periodic solutions of Hill problem because at the very moment he was finishing his paper~\cite{Henon2005}. Later before the thesis defense the second author sent to H\'enon abstract of her thesis and he appreciated of the obtained results.

Planar Hill problem is a limit case of the planar RTBP. The equations of motion of Hill problem describe the motion of massless body near the smaller active mass (the Earth), while the last moves around the bigger active mass (the Sun) along the circular orbit.

Hill problem was originally proposed by American astronomer George Hill for the Moon motion theory.  He had obtained approximate periodic solutions in the form of trigonometric series, which then were used by him as intermediate solutions for his theory of the Moon's motion. For more details about the role of the Hill problem in constructing the theory of the Moon's motion see~\cite{Wilson}.

Hill problem is an nonintegrable Hamiltonian problem and it possesses the infinite number of periodic solution. Some of these solutions can be used for continuation into periodic orbits of the RTBP or into periodic orbits of generic three-body priblem.

M.\,H\'enon made a major contribution to investigations of periodic solutions to planar Hill problem. In papers~\cite{HenonV, HenonVI, Henon74} he summarized the previous results obtained numerically, gave a description of main properties of periodic solutions, and also investigated their vertical stability. 
In papers~\cite{Henon2003,Henon2005} M.\,H\'enon applied the method of generating solutions, used in his book~\cite{Henon97} in the context of planar RTBP, to studying periodic solutions of  planar Hill problem. He also described new families of periodic solutions. In many aspects this paper is a continuation and development of ideas and methods described in~\cite{Henon2003, Henon97} (and also in book~\cite{BrunoRTBPeng} by A.D. Bruno) for seeking periodic solutions based on their generating solutions. However, M.\,H\'enon in the above papers restricted himself to investigation of only families of those periodic solutions which included orbits with a global multiplicity of three and less. Moreover, a major part of families described in paper~\cite{Henon2003} were obtained by him using the ``brute-force'' method. This method turned out to be unsuitable for seeking families of asymmetric solutions. Therefore, all families described in paper~\cite{Henon2005} were found as a result of bifurcation with loss of symmetry of earlier studied families of symmetric periodic solutions.

Hill problem has a lot of applications to analyzing the orbit evolution of natural satellites, to design orbits for space missions, and to studying dynamics of star clusters.

\subsection{Second species generating solutions}
In~\cite{HenonV} H\'enon proposed to look for generating solutions in the set of arc-orbits which start and finish at the origin, which is the only singular point of the Hamiltonian of Hill problem, which is following:
\begin{equation}\label{eqHillHamiltoniancoord}
H(x, y)=\frac12\left(y_1^2+y_2^2\right)+x_2y_1-x_1y_2-x_1^2+\frac12x_2^2-\frac1r,\quad r=\sqrt{x_1^2+x_2^2}.
\end{equation}

There are arc-solutions of two types:
\begin{enumerate}
\item arc-solutions of the first type are arcs of epicycloids, which pass twice through the origin; H\'enon proposed to denote such arcs with symbols $\pm j$, $j\in\mathbb N$;
\item arc-solutions of the second type which are periodic solutions just passing through the origin; such arcs are denoted by symbols $i$ and $e$.
\end{enumerate}

As far as all arc-solutions pass through the origin so it is possible to compose the infinite number of sequences from arc-solutions $\pm j$, $j\in\mathbb N$, $i$, $e$ by matching these arcs with hyperbolas of two types (see~\cite{BatkhinKI2013eng}). Hyperbolas of the first type have their pericenters near $OX$ axis and eccentricity $e\approx1$, hyperbolas of the second type have pericenters near $OY$ axis and eccentricity $e\gg1$.

M.\,H\'enon stated in~\cite{Henon2003} that for  Newtonian potential of attraction there are two pairs of arc-solutions, namely, $ii$ and $ee$, which can not be matched to each other by hyperbolas described above. 

\begin{proposition}[{M.\,H\'enon~\cite{Henon97, Henon2003}}]\label{st1}
A sequence of arc-solutions which does not contain two identical arcs of the second type in succession is a generating solution and it is called \textit{generating sequence} for  the Hill problem.
\end{proposition}

Numerical analysis of all known families of periodic solutions of the  Hill problem allows to state the following
\begin{proposition}\label{st2}
Each family of periodic solutions of Hill problem, which is continuable to solution of the second species, is defined at the limit by generating sequence of Proposition~\ref{st1}.
\end{proposition}

An algorithm for study symmetric periodic solutions defined by its corresponding generating sequences was proposed by the first author in~\cite{Batkhin2012prepr52eng, BatkhinKI2013eng}.
%
More then 20 new families of periodic solutions were found out by this algorithm. Many of them have periodic solutions useful for space flight design~\cite{Batkhin2012prepr52eng, BatkhinKII2013eng}.

\subsection{Connections between Hill problem families}
We applied the approach of generating solutions proposed by M.\,H\'enon for a certain generalization of Hill problem. Let the last term in~\eqref{eqHillHamiltoniancoord} be of the following form: $\sigma/r$, where $\sigma\in\{-1,0,+1\}$, i.\,e. the potential of the central massive body can be a potential of attraction ($\sigma=-1$), can be switched off ($\sigma=0$) or can be a potential of repulsion ($\sigma=+1$). Let call the last case \textit{Anti-Hill problem}. 

\begin{figure}[htb]
\centering
\includegraphics[
width=\linewidth]{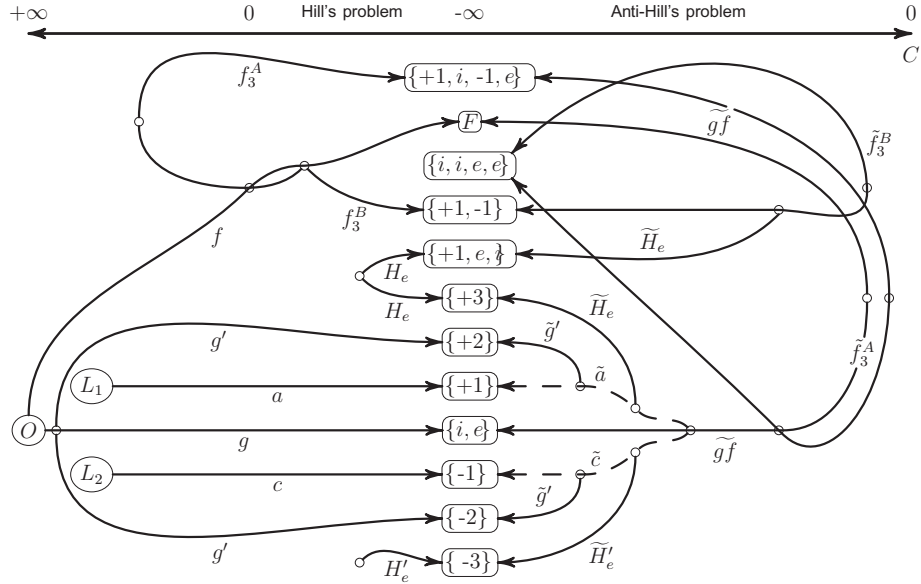}
\caption{Diagram of connection between families of the Hill (left part) and the Ant-Hill problems (right part). Central columns gives generating sequences of the families.}\label{figHillAHilldiagram}
\end{figure}

Figure~\ref{figHillAHilldiagram} gives the result of computations in the form of schematic drawing of families of periodic solutions of Hill and Anti-Hill problems, which form the common network (web) of periodic solutions in the sense that starting from an arbitrary orbit of any family one can continue to any orbit of other family. 

Let give a short description of Figure~\ref{figHillAHilldiagram}. The center column is generating sequences of families. $L_{1,2}$ are libration points, $O$ is the origin. The names of the families are used the same as in M.\,H\'enon's papers~\cite{HenonV,Henon2003}. The linked families of the Anti-Hill problem have the same name with $\tilde{\phantom{w}}$ (tilde) sign above. The common orbits of two families is denoted by small circle. The same circle is used to denote the orbit at which the family riches the extremum on Jacobi integral $C$. To simplify the drawing we show only small part of all known at that moment families. More detail description of connection between families of periodic orbits of Hill and Anti-Hill problem is given in the recent paper~\cite{Batkhin_DAN2014}.

Hill problem like any non integrable problem with closed invariant subset of the phase space  possesses infinite number of periodic solutions grouped into corresponding families. If one would consider Hill problem solely then a lot of families are isolated, i.\,e. they do not share any orbits with other families. We proposed quite natural generalization of Hill problem which demonstrates that all known for us families of periodic orbits of Hill problem connected to each other by both common generating sequences and by sharing common orbits with other families either from Hill problem or from Anti-Hill problem. It means that one can start from any periodic solution and by continuation one can reach any other periodic solution of Hill problem.



\begin{thebibliography}{99}
\bibitem{PoincareI}Poincar\'e~H., Les M\'etods Nouvelles de la M\'ecanique C\'eleste, 1893, Paris~: Gauthier-Villars, Vol.~1.

\bibitem{HenonI}Exploration num\'erique du probl\`eme restreint. {M}asses
  {\'e}gales. {O}rbites periodiques, Annales d'Astrophysique, 1965, Vol.~28, no.~3., P.~499--511.

\bibitem{Henon2005}H\'enon~M., Families of asymmetric periodic orbits in {H}ill's problem
  of three bodies, 2005, Celest. Mech. Dyn. Astr., Vol.~93. , P.~87--100

\bibitem{Henon97}H\'enon~M., Generating Families in the Restricted Three-Body Problem, 1997,
  Lecture Note in Physics. Monographs no.~52. Berlin, Heidelber, New~York~: Springer. 278~p.

\bibitem{HenonII} H\'enon~M., Exploration num\'erique du probl\`eme restreint. {M}asses
  {\'e}gales, stabilit{\'e} des orbites p{\'e}riodiques,1965,  Annales d'Astrophysique., Vol.~28, no.~6, P.~992--1007.

\bibitem{Szebehelyeng} Szebehely~V., Theory of Orbits. The Restricted Problem of Three
  Bodies, 1967, New~York and London~: Academic Press.

\bibitem{Henon68} H\'enon~M., Sur les orbites interplan\'etaires qui rencontrent deux
  fois la Terre, 1968, Bull. Astron., Vol.~3, no.~3., P.~377--402.

\bibitem{BrunoRTBPeng} Bruno~A.~D., The Restricted 3--body Problem: Plane Periodic Orbits, 1994, Berlin~: Walter de Gruyter, 362~p.

\bibitem{HenonV} H\'enon~M., Numerical exploration of the restricted problem. {V}.
  {H}ill's case: Periodic orbits and their stability, 1969, Astron. \&
  Astrophys.,  Vol.~1.,  P.~223--238.

\bibitem{HenonVI} H\'enon~M., Numerical exploration of the restricted problem. {H}ill's
  case: non-periodic orbits, 1970, Astron. \& Astr., no.~9., P.~24--36.

\bibitem{HenonBook2} H\'enon~M., Generating Families in the Restricted Three-Body Problem.
  II. Quantitative Study of Bifurcations, 2001, Lecture Note in Physics. Monographs
  no.~65., Berlin, Heidelber, New~York~: Springer--Verlag, 308~p.

\bibitem{Bruno_stepgeomeng} Bruno~A.~D., Power Geometry in Algebraic and Differential Equations, 2000, Amsterdam~: Elsevier Science, 381~p.

\bibitem{Wilson} Wilson~C., The Hill--Brown Theory of the
  Moon's Motion: Its Coming-to-be and Short-lived Ascendancy (1877-1984).
  Sources and Studies in the History of Mathematics and Physical Sciences, 2010, New York, Dordrecht, Heidelberg, London~: Springer, 323~p.

\bibitem{Henon74} H\'enon~M., Vertical stability of periodic orbits in the restricted
  problem. {II.} {H}ill's case, 1974, Astron. \& Astr., no.~30., P.~317--321.

\bibitem{Henon2003} H\'enon~M., New families of periodic orbits in {H}ill's problem of
  three bodies~, 2003, Celestial Mechanics and Dynamical Astronomy, Vol.~85., P.~223--246.

\bibitem{BatkhinKI2013eng} Batkhin~A.~B., Symmetric periodic solutions of the {Hill's} problem.
  {I}, 2013, Cosmic Research, Vol.~51, no.~4., P.~275--288.

\bibitem{Batkhin2012prepr52eng} Batkhin~A.~B., Symmetric periodic solutions of the Hill's problem, 2012, Preprint No~52., Moscow, Russia~: KIAM,  32~p., in Russian). URL: http://www.keldysh.ru/papers/2012/prep2012\_52.pdf 

\bibitem{BatkhinKII2013eng} Batkhin~A.~B., Symmetric periodic solutions of the {Hill's} problem.
  {II}, 2013,  Cosmic Research, Vol.~51, no.~6., P.~497--510.
  
\bibitem{Batkhin_DAN2014}
Batkhin~A.~B., Web of Families of Periodic Orbits of the Generalized Hill Problem, 2014, Doklady Mathematics, Vol. 90, no.~2, P.~539--544.
	
\end{thebibliography}
\end{document}